\documentstyle[aps,preprint,epsf]{revtex}
 \def\gsim{\mathrel{\rlap{\lower0.2em\hbox{$\sim$}}\raise0.2em\hbox{$>$}}}
 \def\slash{\hskip-0.7em/ \,}
 \def\sumint{\,\int\hspace{-1.3em}\sum\,}
\begin{document}

\title{Consistent HTL resummation of the thermodynamical potential}

\author{Andr\'e Peshier\footnote{present address:
        Brookhaven National Laboratory, Upton, New York 11973-5000, USA}}
\address{Forschungszentrum Rossendorf, PF 510119, 01314 Dresden, Germany,
         \\
         Institut f\"ur Theoretische Physik, Technische Universit\"at
         Dresden, 01062 Dresden, Germany}

\maketitle

\begin{abstract}
 The thermodynamical potential of relativistic plasmas with gauge
 interaction can be consistently resummed in terms of HTL propagators,
 which is, without being restricted to it, exemplified for the case of
 hot QED.
 The nonperturbative resummation obtained in a $\Phi$-derivable approach is
 gauge independent, free of thermal divergences and, in the weak-coupling
 limit, compatible with the leading order perturbative result.
\end{abstract}


\section{Introduction}
 One of the main issues of the ongoing heavy-ion program is the
 investigation of deconfined hadronic matter. Despite asymptotic freedom
 of the strong interaction, this quark-gluon plasma is characterized by
 a large coupling in the regimes of physical interest, so nonperturbative
 approaches are required to describe this many-particle system reliably.

 Recently, within the framework of an equilibrium description of the
 QCD plasma, the calculation of thermodynamical quantities by resumming
 hard-thermal-loop (HTL) propagators was proposed.
 In \cite{ABS} and \cite{BR}, nonperturbative expressions were given
 for the thermodynamical potential which, however, are detracted from
 inconsistencies: The expressions do not reproduce, in the weak-coupling
 limit, the perturbative results at leading order ${\cal O}(g^2)$ and they
 are, with uncompensated medium-dependent divergences, not satisfactory from
 a formal point of view. These problems were claimed to be solved only in a
 fully resummed calculation.
 In Ref.\ \cite{BIR1}, instead, a consistent approximation of the entropy
 was derived from HTL propagators. This approach in principle resolves the
 problem of a leading order thermodynamical resummation since, up to an
 integration constant, the thermodynamical potential can be reconstructed
 from the entropy \cite{BIR2}.

 Nonetheless, a direct calculation of the thermodynamical potential, the
 quantity containing the full thermodynamical information, is instructive,
 in particular since an approach to derive macroscopic properties, which
 are sensitive to hard momenta, from HTL propagators (as a soft-momentum
 approximation) is, a priori, far from obvious.
 On the other hand, being explicitly gauge independent and respecting for
 arbitrary momenta the fundamental sum rules resulting from the commutator
 relations, the HTL propagators are a favorable basis for calculating
 physical quantities from approximate dressed Green's functions, yielding
 (mostly) analytical nonperturbative results.
 These points are addressed in the present note where, starting from the
 Luttinger-Ward representation of the thermodynamical potential \cite{LW},
 the hot QED plasma is studied.
 This case is particularly simple to analyze but at the same time also
 representative for systems with an equal HTL structure as, e.\,g., the
 quark-gluon plasma at finite temperature and density.

\section{The approximation scheme}
 As an exact relation, the thermodynamical potential can be expressed
 in terms of fully dressed Green's functions by the (generalized)
 Luttinger-Ward representation \cite{LW,DM,VB}
 \begin{equation}
   \Omega
   =
   \frac12\, \tilde\Omega[D] - \tilde\Omega[S] + \Phi[D,S] \, .
  \label{Omega}
 \end{equation}
 $\tilde\Omega$ is a functional of the photon propagator $D$ or the electron
 propagator $S$, which are related by Dyson's equation to the respective
 self-energies. The boson part, e.\,g., is defined by
 \[
   \frac12\, \tilde\Omega[D]
   =
   \frac12\, \mbox{Tr}\,[ \ln(-D^{-1}) + D\Pi ] \, ,  \quad
   D^{-1} = D_0^{-1} - \Pi \, ,
 \]
 where the trace is taken over the four-momentum and the Lorentz
 structure, while in the analog fermion part $-\tilde\Omega[S]$,
 with $S^{-1} = S_0^{-1} - \Sigma$, the spinor indices are traced.
 The functional $\Phi[D,S]$ given by all two-particle irreducible bubble
 graphs with exact propagators (`dressed skeletons') is related to the
 self-energies by
 \begin{equation}
   \Pi = -2\, \frac{\delta\Phi}{\delta D} \, , \quad
   \Sigma = \frac{\delta\Phi}{\delta S} \, .
  \label{SE}
 \end{equation}
 Consequently, the fundamental stationarity of the thermodynamical potential
 upon variation of the self-energies \cite{LY} is fulfilled by the
 representation (\ref{Omega}). It is emphasized that, on account of its
 stationarity, $\Omega$ is formally gauge independent, although the
 propagators are not.

 Using the projectors ${\cal P}_{\mu\nu}^L = -\tilde{K}_\mu\tilde{K}_\nu /K^2$
 and ${\cal P}_{\mu\nu}^T = g_{\mu\nu}-K_\mu K_\nu /K^2-{\cal P}_{\mu\nu}^L$,
 where $\tilde{K} = [K(Ku)-uK^2]/[(Ku)^2-K^2]^{1/2}$ and $u$ is the medium
 four-velocity, the inverse photon propagator is decomposed into the
 transverse ($T$) and the longitudinal ($L$) part as well as the covariant
 gauge-fixing term,
 \[
   D_{\mu\nu}^{-1}(K)
   =
   \sum_{i=T,L} {\cal P}_{\mu\nu}^i \Delta_i^{-1} + \frac1\xi K_\mu K_\nu \, ,
   \quad
   \Delta_i^{-1} = \Delta_0^{-1} - \Pi_i \, ,
 \]
 with $\Delta_0^{-1} = K^2 = k_0^2-k^2$. Introducing the fermionic
 `projectors' ${\cal P}_\pm(K) = \frac12\, (K\slash \pm \tilde{K}\slash)$
 on the particle and the hole excitations (the index denotes the ratio of
 chirality to helicity), the electron propagator can be written in a similar
 way as
 \[
   S = \sum_{i=\pm} {\cal P}_i \Delta_i \, , \quad
   \Delta_i^{-1} = \Delta_0^{-1} - \Sigma_i \, .
 \]
 In terms of the scalar propagators $\Delta_i$, with the degeneracy
 factors $d_T=d-1$, $d_L=1$ and $d_\pm=(d+1)/2$, the $\tilde\Omega$
 parts of eq.~(\ref{Omega}) read
 \begin{eqnarray}
   \frac12\, \tilde\Omega[D]
   &=&
   \frac12\, \sumint\left\{
      \sum_{i=T,L} d_i \left[ \ln(-\Delta_i^{-1})+\Delta_i\Pi_i \right]
    - \ln(-\Delta_0^{-1})
   \right\} ,
   \nonumber \\
   \tilde\Omega[S]
   &=&
   \sumint\left\{
      \sum_{i=\pm} d_i \left[
         \ln(-\Delta_i^{-1})+\Delta_i\Delta_0^{-1} \right]
     -d_\pm\ln(-\Delta_0^{-1})
   \right\} ,
  \label{OmegaTilde}
 \end{eqnarray}
 where the subtractive contribution of the ghost fields, which otherwise
 decouple, is included in $\tilde\Omega[D]$, and the integral-sums,
 continued to $d = 3 - 2\varepsilon$ spatial dimensions,
 \[
   \sumint = \int_{k^d} T\sum_{k_0} \, , \quad
   \int_{k^d} = \mu^{2\varepsilon} \int\frac{d^d k}{(2\pi)^d} \, ,
 \]
 run over either bosonic or fermionic Matsubara frequencies $k_0$.

 Commencing from this exact approach, selfconsistent (`symmetry conserving')
 approximations can be derived \cite{Ba}: an approximation of the functional
 $\Phi$ in the scheme (\ref{Omega}-\ref{OmegaTilde}) yields approximate
 self-energies and an expression for the thermodynamical potential which is
 still stationary.
 In particular, the loop expansion of $\Phi$ can be truncated at a certain
 order, which is, in terms of the perturbative expansion in free Green's
 functions, equivalent to a partial resummation avoiding the problem of
 double counting of diagrams.
 The leading-loop order of the $\Phi$-derivable approximation is given
 diagrammatically by\footnote{It is noted that the (vacuum) counter terms
   required to render the self-energies finite can be implemented in the
   $\Phi$-derivable scheme by appropriate {\em counter loops} in $\Phi$,
   leaving $\Omega$ itself unchanged.}
 \[
   \Phi^{ll} = -\frac12 \epsfysize=5mm \epsffile{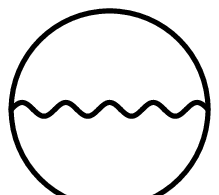} \, , \quad
   \Pi^{ll}  = \epsfysize=5mm \epsffile{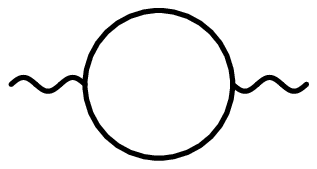} \, , \quad
   \Sigma^{ll} = - \epsfysize=5mm \epsffile{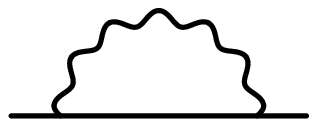} \, ,
 \]
 which implies the relation
 \begin{equation}
   \Phi^{ll}
    =
    -\frac12\, \mbox{Tr} D^{ll} \Pi^{ll}
    =
    \frac12\, \mbox{Tr} S^{ll} \Sigma^{ll}
   \label{PhiLL}
 \end{equation}
 among the selfconsistent solutions of the coupled Dyson equations and
 $\Phi^{ll}$.

 In the following, the selfconsistent one-loop self-energies are approximated
 by the HTL self-energies $\Pi^\star$ and $\Sigma^\star$.
 Although the HTL self-energies remain a reasonable approximation even for
 hard momenta \cite{KKR,PST}, it is not evident a priori that a consistent
 approximation of $\Omega$ can be formulated in terms of such approximated
 quantities which are derived for soft momenta much smaller than the
 temperature $T$, whereas thermodynamics is sensitive to the momentum scale
 $T$. In fact, the HTL approximation of the self-energies undermines the
 selfconsistency of the $\Phi$-derivable approach since the relation analogous
 to (\ref{PhiLL}) is not fulfilled.
 Nevertheless, as shown in the following, it is possible to consistently
 resum the HTL contributions to the thermodynamical potential, yielding a
 nonperturbative `continuation' of the ${\cal O}(e^2)$ perturbative result.
 In Sec.~III, the HTL contributions to the $\tilde\Omega$ parts of the
 thermodynamical potential are calculated directly from the HTL self-energies
 since, e.\,g., replacing $\Pi^{ll}$ by $\Pi^\star$ is correct up to terms
 beyond the order ${\cal O}(e^2)$ under consideration.
 The corresponding $\Phi$ contribution is then given in Sec.~IV to complete
 the resummed approximation of the thermodynamical potential which is,
 expressed by HTL self-energies, explicitly gauge invariant.

\section{The \boldmath{$\tilde\Omega$} contributions}
 The HTL self-energies of the photon and the electron are given by \cite{lB}
 \begin{eqnarray}
   \Pi_T^\star
   &\!=\!&
   M_b^2+\tilde\Pi \, , \quad
   \Pi_L^\star
   \,=\,
   -2\tilde\Pi \, , \quad
   \tilde\Pi(k_0,k)
   =
   M_b^2\, \frac{K^2}{k^2}
   \left[ 1+\frac{k_0}{2k}\ln\frac{k_0-k}{k_0+k} \right] ,
   \nonumber \\
   \Sigma_\pm^\star
   &=&
   \frac12\, M_f^2 \pm \tilde\Sigma \, , \quad
   \tilde\Sigma(k_0,k)
   =
   \frac{M_f^2}2
   \left[ \frac{k_0}{k} + \frac{K^2}{2k^2}\ln\frac{k_0-k}{k_0+k} \right] .
  \label{SE-HTL}
 \end{eqnarray}
 The quantities
 \begin{equation}
   M_b^2 = \frac{e^2 T^2}6 \, , \quad
   M_f^2 = \frac{e^2 T^2}4
  \label{asympt_mass}
 \end{equation}
 can be considered as asymptotic masses (squared) of the transverse photon
 and the electron particle excitation, respectively, since their dispersion
 relations approach mass shells for momenta $k \gg eT$.
 The longitudinal photon (plasmon) mode and the hole (plasmino) excitation,
 on the other hand, possess a vanishing spectral strength when approaching
 the light cone exponentially fast for $k \gsim eT$.

 Besides the $\Phi$ part, eq.~(\ref{OmegaTilde}) evaluated with HTL
 propagators yields the contributions to the thermodynamical potential
 which are, up to gauge group factors, the same as for non-Abelian gauge
 theories with HTL self-energies with a structure like (\ref{SE-HTL}).
 In Refs.\ \cite{ABS,BR}, however, where the QCD plasma was studied in
 the framework of the leading order HTL perturbation theory, only the
 $\ln(-\Delta_i^{-1})$ terms of (\ref{OmegaTilde}) were considered to
 contribute to the thermodynamical potential.
 In terms of the bare-propagator expansion, this incomplete analysis amounts to
 a miscounting of graphs, so in the weak-coupling expansion the leading order
 perturbative contribution to the thermodynamical potential is not reproduced
 correctly in either the case of hot QCD with vanishing chemical potential
 \cite{ABS} or for the degenerate quark-gluon plasma \cite{BR} at $T=0$.
 For the QED plasma under consideration, the corresponding boson ($b$) and
 fermion ($f$) contributions, marked here by the index $A$, follow from the
 expressions given in \cite{ABS} by replacing the asymptotic gluon and quark
 masses by the expressions (\ref{asympt_mass}), schematically
 \begin{eqnarray}
   \tilde\Omega_{A,b}^\star
   &=&
   \frac12\, \sumint \left\{
     d_T \ln(-\Delta_T^{\!\star\;-1})
    + \ln(-\Delta_L^{\!\star\;-1}) -\ln(-\Delta_0^{-1}
   \right\}
   =
   -\frac{M_b^4}{32\pi^2} \frac1\varepsilon +  \mbox{finite terms} \, ,
   \nonumber \\
   \tilde\Omega_{A,f}^\star
   &=&
    \sumint d_\pm \left\{
      \ln(-\Delta_+^{\!\star\;-1})
     +\ln(-\Delta_-^{\!\star\;-1})-\ln(-\Delta_0^{-1})
    \right\}
    =
    \mbox{finite terms} \, .
   \label{partA}
 \end{eqnarray}
 The imperfect cancelation of the thermal divergences in
 $\tilde\Omega_{A,b}^\star$ is a second, formal indication of missing
 contributions in the approaches \cite{ABS,BR}.

 The complete analysis of the HTL-resummed thermodynamical potential has
 to keep track of the remaining terms, indexed $B$ in the following, of
 eq.~(\ref{OmegaTilde}). Using complex contour integration, the Matsubara
 sums can be calculated to yield a quasiparticle part stemming from the pole
 $\omega_i(k)$ of the propagators, and a Landau-damping contribution arising
 from the discontinuity of the HTL self-energies below the light cone,
 \[
   \sum_{k_0} \Delta_i^\star f_i
   =
   -(1 + 2n(\omega/T))
    \left.
      \frac{f_i}{\partial_\omega \Delta_i^{\!\star\;-1}}
    \right|_{\omega_i}
   +\int_0^k \frac{d\omega}{2\pi}\, (1 + 2n(\omega/T)) \Psi_i \, , \quad
   \Psi_i = \mbox{Disc}(\Delta_i^{\! \star} f_i) \, ,
 \]
 where $n = \pm n_{b,f}$ is either the Bose or the negative of the
 Fermi distribution function, and $f_i$ can be $\Pi_{T,L}^\star$ or
 $\Delta_0^{-1}$.
 Subtracting and adding the appropriately infrared-regularized asymptotic
 integrands analog to the technique applied in \cite{ABS}, the integrals
 over the spatial momenta are split into a finite contribution and the
 dimensionally regularized subtraction term.
 Taking the limit $\varepsilon \rightarrow 0$, the individual contributions
 are
 \begin{eqnarray}
   \tilde\Omega_{B,b}^\star
   &=&
   \frac12 \sum_{i=T,L} \sumint d_i\, \Delta_i^{\!\star} \Pi_i^\star
   =
   \frac12 \sum_{i=T,L} d_i
   \int_{k^3} \left[
      \left.
        \frac{-(1+2n_b) \Pi_i^\star}{2\omega-\partial_\omega \Pi_i^\star}
      \right|_{\omega_i}
     +\int_0^k \frac{d\omega}{2\pi}\, (1+2n_b) \Psi_i
     -B_i^{sub}
   \right]
   \nonumber \\
   && \hskip 3cm
   + \frac{M_b^4}{32\pi^2}
      \left[
        \frac2\varepsilon
       +2\ln\frac{4\pi}{e^\gamma}
       +\frac13 \left( \frac{14}3-2\pi^2+\frac{16}3\,\ln2-8\ln^2 2 \right)
      \right] \left( \frac{M_b^2}{\mu^2} \right)^{\! -\varepsilon} \!\!,
   \nonumber \\
   \tilde\Omega_{B,f}^\star
   &=&
   \sum_{i=\pm} d_i  \sumint \Delta_i^{\!\star} \Delta_0^{-1}
   =
   d_\pm \sum_{i=\pm}
    \int_{k^3}  \left[
           \left.
              \frac{-(1-2n_f) K^2}{2\omega-\partial_\omega \Sigma_i^\star}
           \right|_{\omega_i}
          + \int_0^k \frac{d\omega}{2\pi}\, (1-2n_f) \Psi_i
          -B_i^{sub}
        \right]
   \nonumber \\
   && \hskip 3cm
   +\frac{M_f^4}{8\pi^2}\, \ln2\, ( -1+2\ln2 ) \, ,
  \label{partB}
 \end{eqnarray}
 with the angles
 $\Psi_{T,L} = \mbox{Disc}(\Delta_{T,L}^{\!\star} \Pi_{T,L}^\star)$,
 $\Psi_\pm = \mbox{Disc}(\Delta_\pm^{\!\star} \Delta_0^{-1})$,
 and the subtraction terms
 \begin{eqnarray*}
   \sum_{i=T,L} d_i B_i^{sub}
   &=&
   -d_T \left(
     \frac{M_b^2}{2k}
    +\frac{M_b^4}{2k(k^2+M_b^2)}
      \left( 2-\ln\frac{4(k^2+M_b^2)}{M_b^2} \right)
  \right)
  \\
  && \hskip -3mm
  -\int_0^k \frac{d\omega}\pi\, \mbox{Im}\tilde\Pi
     \left[
       \frac{-d_T K^2}{(K^2-M_b^2)^2}
         \left( 1+2\frac{\mbox{Re}\tilde\Pi}{K^2-M_b^2} \right)
      +\frac{2}{K^2}
         \left(
           1-4\,\frac{\mbox{Re}\tilde\Pi}{K^2}\frac{k^2}{k^2+M_b^2}
         \right)
     \right] ,
  \\
   \sum_{i=\pm} B_l^{sub}
   &=&
   - \frac{M_f^2}{2k}
   - \frac{M_f^4}{8k(k^2+M_f^2)} \left( 1-2\ln\frac{4(k^2+M_f^2)}{M_f^2} \right)
   \\
   &&
   - \int_0^k \frac{d\omega}\pi\, \mbox{Im}\tilde\Sigma
           \left[ \frac1{K^2} - \frac{K^2}{K^2-M_f^2} \right] .
 \end{eqnarray*}
 It turns out in eq.~(\ref{partB}) that the thermal divergences of the
 fermion integral-sum cancel, as for $\tilde\Omega_{A,f}^\star$ in
 (\ref{partA}).
 The boson part, on the other hand, contains a temperature dependent term
 $\sim M_b^4/\varepsilon$ which is twice as large in magnitude as its
 counterpart in (\ref{partA}).
 These thermal divergences of the $A$ and $B$ contribution have to cancel the
 corresponding terms in the HTL approximation of $\Phi$, which is calculated
 in the following section, to yield a well-defined resummation of the
 thermodynamical potential.

\section{The \boldmath{$\Phi$} contribution}
 In this section, the remaining $\Phi$ contribution is calculated by
 evaluating the two-loop functional, given the self-energies and the
 Green's functions.\footnote{In the approach \cite{BIR1,BIR2}, instead, the
     functional properties of $\Phi$ in the leading loop approximation are
     used first to derive the entropy, which is then evaluated with the HTL
     propagators. It is noted, however, that in this framework the
     selfconsistency relation only holds approximately, see below.}
 Within the leading loop approximation, $\Phi$ is related to the self-energies
 by equation (\ref{PhiLL}) and can thus be expressed in a general form by
 $\Phi^{ll} = -\frac12\, t\,$Tr$D^{ll}\Pi^{ll} + \frac12\, (1-t)\, $Tr$S^{ll}
 \Sigma^{ll}$, independently of $t$.
 In the framework of the HTL approximation, however, the consistency relation
 (\ref{PhiLL}) is violated since the traces are dominated by hard momenta.
 The resulting ambiguity is even of order ${\cal O}(e^2)$, namely
 \[
   \left. \epsfysize=5mm \epsffile{phi.eps}\!\!\right|_{\rm lo}
   =
   \,\left. \frac54\epsfysize=5mm \epsffile{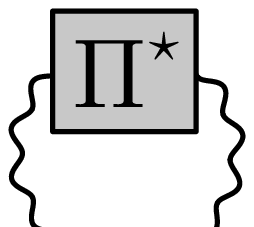} \right|_{\rm lo}
   =
   \,\left. \frac53\epsfysize=5mm \epsffile{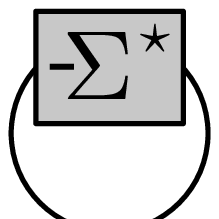}\!\!\right|_{\rm lo} .
 \]
 Hence, the HTL contribution to $\Phi$ cannot be obtained by naively replacing
 the leading loop quantities in $\Phi^{ll}$, as given above, by their HTL
 approximation.
 Instead, the HTL contribution $\Phi^\star$ can be conceived by analyzing how
 the ${\cal O}(e^2)$ discrepancy arises.
 Denoting the photon momentum in \raise0.7mm\hbox{\epsfysize2.2mm
 \epsffile{phi.eps}}\ by $K$ and the fermion momenta by $Q_{1,2}$, this
 diagram (with bare propagators for the ${\cal O}(e^2)$ contribution) can be
 represented as a double integral-sum over an expression with a numerator
 $N = K^2-Q_1^2-Q_2^2$.
 Closing the external legs of the boson self-energy in the HTL approximation
 amounts to neglecting the term $K^2$ in $N$. Tracing over the negative of the
 fermion HTL self-energy, on the other hand, neglects one of the $Q^2$ terms
 in $N$. Thus, all terms are accounted for twice in the sum over all three
 possibilities to approximate one of the momenta as soft.
 Accordingly, the representation
 \begin{equation}
   \Phi^\star
   =
   -\frac14\, \mbox{Tr}D^\star \Pi^\star
   +\frac12\, \mbox{Tr}S^\star \Sigma^\star
  \label{PhiStar}
 \end{equation}
 reproduces $\Phi^{ll}$ perturbatively to order ${\cal O}(e^2)$.
 As shown in the following, this representation indeed leads to a well-defined
 resummed approximation of the thermodynamical potential.

 It is first emphasized that in the complete expression resulting from the
 HTL approximation of eq.~(\ref{OmegaTilde}) and (\ref{PhiStar}), which can
 be written in a compact form as
 \begin{equation}
   \Omega^\star
   =
   \frac12\, \mbox{Tr} \left[
     \ln(-D^{\star\,\, -1}) + \frac12\, D^\star \Pi^\star
   \right]
   - \mbox{Tr} \left[
     \ln(-S^{\star\,\, -1}) + \frac12\, S^\star \Sigma^\star
   \right]
   - \Omega_{\rm ghost} \, ,
  \label{OmegaStar}
 \end{equation}
 with the individual contributions given by (\ref{partA}), (\ref{partB}),
 all temperature dependent divergences cancel.
 Moreover, the perturbative limit of the thermodynamical potential is
 reproduced by the representation (\ref{OmegaStar}). Separating the
 free contributions, e.\,g., for the boson part by $\ln(-D^{-1}) =
 \ln(-D_0^{-1})+\ln(1-D_0\Pi)$, and using the expansion $\ln(1-x)+x/2 =
 -x/2+{\cal O}(x^2)$, the leading order correction to the interaction-free
 limit $\Omega_0 = -(d_T+\frac78\,2d_\pm)\, \frac{\pi^2}{90}\, VT^4$ is
 \begin{equation}
   \Omega_{\rm lo}^\star
   =
   -\frac14\, \mbox{Tr} D_0 \Pi^\star
   +\frac12\, \mbox{Tr} S_0 \Sigma^\star
   =
   \frac{T^2}{24}\, (M_b^2+M_f^2)
   =
   \Omega_{\rm lo}^{pert} \, .
 \end{equation}
 As in the HTL calculations \cite{BIR1,BIR2} of the QCD entropy, the leading
 order term originates entirely from the behavior of the thermodynamically
 relevant excitations at the hard momentum scale $T$, which {\em a posteriori}
 justifies the present approach.
 However, in contrast to the entropy calculations \cite{BIR1,BIR2} where the
 results are manifestly ultraviolet-finite, the cancelation of the thermal
 divergences is in the present approach directly related to the fact that the
 perturbative result is reproduced.
 It is emphasized that this aspect makes the representation (\ref{OmegaStar})
 of $\Omega^\star$ unique; any linear combination, apart from (\ref{PhiStar}),
 of closed self-energy diagrams for the $\Phi$-contribution would result in
 either uncompensated thermal divergences or an incorrect perturbative limit.

 The next-to-leading order term of the perturbative result $\Omega^{pert}$,
 on the other hand, cannot be expected to be reproduced in the present
 approximation: In an equivalent approach for the scalar $g^2 \phi^4$
 theory, where the complete leading loop Luttinger-Ward resummation of the
 thermodynamical potential can be derived \cite{PKS}, the corresponding
 ${\cal O}(g^3)$ correction takes its correct value only after the
 resummation of the self-energy, while the expressions (\ref{SE-HTL}) are
 calculated with bare propagators.
 Accordingly, the contribution of order ${\cal O}(e^3)$, which arises from
 the static longitudinal parts of (\ref{OmegaStar}),
 \[
   \Omega_{\rm nlo}^\star
   =
   \left. \Omega_{A,b}^\star\right|_{\rm nlo}
   +\frac12\, \left. \Omega_{B,b}^\star\right|_{\rm nlo}
   =
   -\frac{(2M_b^2)^{3/2}\, T}{12\pi} + \frac{(2M_b^2)^{3/2}\, T}{16\pi} \,
 \]
 is found to underestimate the perturbative result $\Omega_{\rm nlo}^{pert}
 = -(e^2 T^2/3)^{3/2}\, T/(12\pi)$ by a factor of 1/4.
 As observed in \cite{ABS,BR} for QCD, the next-to-leading order term
 of $\Omega_{A,b}^\star$ agrees with the perturbative result but is 
 overcompensated by the $\Omega_{B,b}^\star$ contribution, which is to 
 be interpreted by a systematic next-to-leading order calculation.

\section{Summary}
 The generalized Luttinger-Ward representation of the thermodynamical
 potential is a suitable framework to derive consistently resummed
 approximations for relativistic gauge theories, with the propagators
 approximated by their HTL contributions.
 While the formalism is particularly simple to analyze in the case of
 the hot QED plasma, which is exemplified here, the application to
 other systems (including the QCD plasma) with the same HTL structure
 is evident.
 As a direct result of the HTL approximation, the resummed thermodynamical
 potential (\ref{OmegaStar}) is gauge independent.
 All medium-dependent divergences cancel, hence the approximation is
 explicitly renormalization-scale independent.
 The resummed expression (\ref{OmegaStar}) enjoys the anticipated behavior
 of a nonperturbative approximation.
 As shown in Figure 1, it yields a smooth extrapolation to the
 large-coupling regime, where it is inclosed by the perturbative results
 which are known to fluctuate with increasing order. In the weak-coupling
 limit, on the other hand, the perturbative result is recovered.
 This demonstrates that to leading order the resummed thermodynamical
 potential can entirely be expressed in terms of HTL propagators.

 The applications of the formalism to the hot and dense quark-gluon plasma
 are straightforward and promising, in particular with regard to the
 suggestion \cite{PKS2} to extrapolate finite-temperature lattice data to
 non-zero chemical potential.
 The reliability of the leading order HTL resummation of the thermodynamical
 potential in the large coupling regime, however, remains to be justified by
 a systematic next-to-leading order calculation.
 \\[3mm]
 {\bf Acknowledgments:} I thank J.~Knoll and F.~Gelis for helpful discussion
 and A.~Rebhan for useful comments on the manuscript.

 \begin{figure}[h]
  \centerline{\epsfxsize=12cm \epsffile{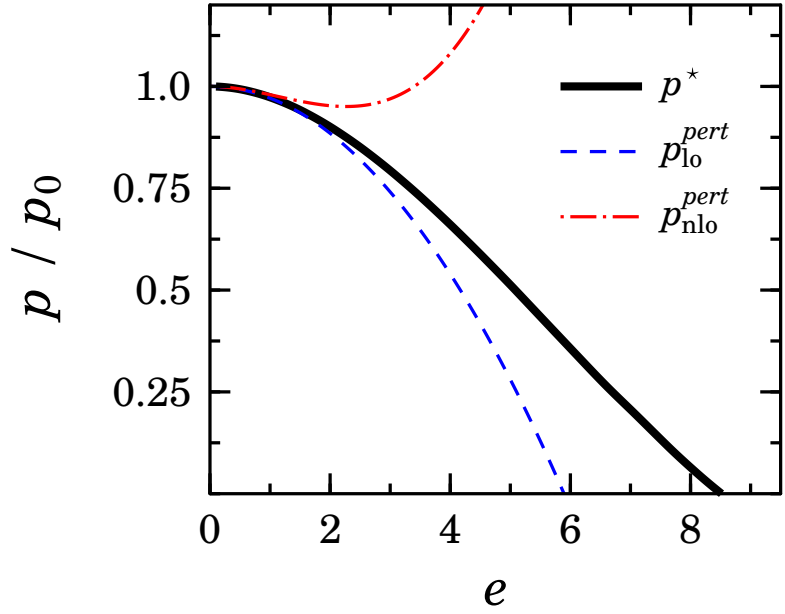}}
  \caption{The resummed pressure $p^\star = -\Omega^\star/V$ (solid line),
           as a function of the coupling and scaled by the interaction-free
           limit $p_0=(2+\frac78\,4)\,\frac{\pi^2}{90}\, T^4$, compared to
           the perturbative result in leading order (dashed line) and in
           next-to-leading order (dash-dotted line).}
 \end{figure}


\begin{references}
 \bibitem{ABS} J.\,O.\ Andersen, E.\ Braaten, M.\ Strickland,
              Phys.\ Rev.\ Lett.\ 83 (1999) 2139,
              Phys.\ Rev.\ D61 (2000) 014017, hep-ph/9908323
 \bibitem{BR} R.\ Baier, K.\ Redlich,
              Phys.\ Rev.\ Lett.\ 84 (2000) 2100
 \bibitem{BIR1} J.\,P. Blaizot, E.\ Iancu, A.\ Rebhan,
              Phys.\ Rev.\ Lett.\ 83 (1999) 2906
 \bibitem{BIR2} J.\,P. Blaizot, E.\ Iancu, A.\ Rebhan,
              Phys.\ Lett.\ B470 (1999) 181
 \bibitem{LW} J.\,M.\ Luttinger, J.\,C.\ Ward,
              Phys.\ Rev.\ 118, (1960) 1417
 \bibitem{DM} C.\ De\,Dominicis, P\,.C.\ Martin,
              J.\ Math.\ Phys.\ 5 (1964) 14, 31
 \bibitem{VB} B.\ Vanderheyden, G.\ Baym,
              J.\ Stat.\ Phys.\ 93 (1998) 843
 \bibitem{LY} D.\,T.\ Lee, C.\,N.\ Yang,
              Phys.\ Rev.\ 117 (1960) 22
 \bibitem{Ba} G.\ Baym,
              Phys.\, Rev.\ 127 (1962) 1391
 \bibitem{KKR} U.\ Kraemmer, M.\ Kreutzer, A.\ Rebhan,
              Ann.\ Phys.\ 201 (1990) 223
 \bibitem{PST} A.\ Peshier, K.\ Schertler, M.\,H.\ Thoma,
              Ann.\ Phys.\ 266 (1998) 162
 \bibitem{lB} M.\ le\,Bellac,
              {\em Thermal Field Theory}, Cambridge University Press (1996)
 \bibitem{PKS} A.\ Peshier, B.\ K\"ampfer, G.\ Soff,
              Euro.\ Phys.\ Lett.\ 43 (1998) 381
 \bibitem{PKS2} A.\ Peshier, B.\ K\"ampfer, G.\ Soff,
              hep-ph/9906305
\end{references}
\end{document}